\documentclass[manuscript]{aastex}

\shorttitle{Magnetic Fields and Velocity Structures around a Cancellation Site}
\shortauthors{Iida et al.}

\begin{document}

\title{Vector magnetic fields and Doppler velocity structures \\
 around a cancellation site in the Quiet-Sun}

\author{Y. Iida and T. Yokoyama}
\affil{Department of Earth and Planetary Science, University of Tokyo, Hongo, Bunkyo-ku, Tokyo, 113-0033, Japan}
\email{iida@eps.s.u-tokyo.ac.jp}
    
\and

\author{K. Ichimoto}
\affil{Hida Observatory, Kurabashira Kamitakara-cho, Takayama-city, Gifu, 506-1314, Japan}

\begin{abstract}
A cancellation is thought to be a basic process of the photospheric magnetic field and plays an important role
 in magnetic flux budget and in various solar activities.
There are two major theoretical scenarios for this phenomena, i.e. the ``U-loop emergence'' and the ``$\Omega$-loop submergence'' models. 
It is important to clarify which is the dominant process during the cancellation for the estimation of the solar magnetic
 flux transport through the surface.
We study the vector magnetic field and velocity structures around a quiet Sun cancellation by using the Solar Optical
 Telescope on board Hinode satellite.
Transverse magnetic field connecting the canceling magnetic features and strong long-lasting Doppler red-shift signal are found.
The transverse field is observed in the first spectropolarimetric observation after the occurrence of the cancellation while the red-shift clearly delayed to the cancellation by 20 minutes.
These results indicate that the observed cancellation is an ``$\Omega$-loop submergence''.

\end{abstract}


\section{Introduction}

 A magnetic cancellation is an apparent convergence with a following disappearance of positive 
and negative magnetic patches observed in the line-of-sight (LOS) magnetograms of the solar photosphere \citep{liv85,mar85}.
 A cancellation is thought to be important not only for the photospheric magnetic flux budget \citep{sch97} but also
 for various solar activities (e.g. solar flares, X-ray bright points, and filament formation). 
 The physical structure and the dynamics of this phenomenon have, however, not been understood yet.
 Two theoretical scenarios are proposed.
 One is a ``U-loop emergence'' scenario and the other is an ``$\Omega$-loop submergence'' scenario \citep{zwa87}.
 Because these two scenarios are totally different in the flux transport between below the photosphere and the
 solar atmosphere, it is important to distinguish these by the observations.

 The first observational study on the cancellation scenarios is reported by Harvey et al. (1999). 
 They investigated whether the magnetic loop submerges or emerges by determining the temporal evolution of
 the photospheric and chromospheric magnetic fields.
 If a cancellation occurs at the chromosphere after at the photosphere, it is interpreted as an emergence of a flux
and vise versa.
 They found that a flux submergence is the most frequent (44$\%$) as compared with a flux emergence (18$\%$) and with other
 38$\%$ that shows no significant time difference.
 They also derived the submergence velocity ($\thicksim 1.0 \ {\rm km}\ {\rm s}^{-1}$) by dividing the relative height
 between the two observed layers by the time difference.
 Chae et al. (2004) succeeded in the direct measurement of the Doppler velocity structure around
 cancellation sites by using the spectroscopic analysis.
 They found a strong horizontal field and downflow ($\thicksim 1.0 \ {\rm km}\ {\rm s}^{-1}$) around two cancellation sites 
in an active region.
 Yang et al. (2009) also found the similar downflow ($\thicksim 0.9 \ {\rm km}\ {\rm s}^{-1}$) at the cancellation sites in
 a coronal hole.
 These signatures support the ``$\Omega$-loop submergence'' scenario.
 
 On the contrary, there are some papers reporting the blue-shifts and no horizontal field around the cancellation sites.
 Zhang et al. (2001) reported that they could not observe the transverse field connecting the canceling patch around the 
cancellations related with the filament eruption (also see Wang \& Shi(1993)). 
 Kubo $\&$ Shimizu (2007) investigated the vector magnetic field and the Doppler velocity field around 14 cancellations
 in active regions.
 The result is that there are both upflow and downflow.
 They also found a center-to-limb variation that the absolute value of the observed velocity becomes larger near the
 limb than on the disk center.
 They concluded that these velocity structures indicate the material flow along the horizontal field.
 Zhang et al. (2009) investigated the interaction between the magnetic activities (e.g. flux emergences and cancellations) and 
the granular motion and reported the upflow at the cancellation site.
 Their interpretation of their cancellations is a ``U-loop emergence''.
 The velocity field and the horizontal field around the cancellation sites are still open questions.

 In this paper, we study the vector magnetic field and velocity structures of a cancellation in a quiet Sun by using
 high-temporal and high-spatial resolution observations of the Solar Optical Telescope on board Hinode.
 The purpose of this study is to distinguish whether a cancellation is a ``U-loop emergence'' or an ``$\Omega$-loop submergence''.

\section{Observations and Analysis}

 We use the data taken by the Solar Optical Telescope (SOT) onboard the Hinode spacecraft in this study
 \citep{kos07,tsu08,ich08,sue08,shi08}.
 The image sequence obtained by the SOT Narrowband Filter Imager (NFI) of the Filtergram (FG) on 2007 October 31th is used. 
 The region is a quiet Sun in the south east part of the disk; the center of the field of view is located at
 ($ -500\ {\rm arcsec} , -200\ {\rm arcsec}$) in the helio centric coordinate.
 Stokes I and V components positioned at $\pm 156\ {\rm m}\AA$ of the Na I 5896 $\AA$ line at the observed wavelength
 are obtained by the NFI
 with pixel-resolution of 0.16 arcsec with an 82 arcsec $\times$ 82 arcsec field of view.
 The cadence is 1 minute and the resolution in wavelength is  0.1 $\AA$.
 The contribution of NaI 5896 $\AA$ line is the upper photosphere and the lower the chromosphere.
 We determine the temporal evolution of the  LOS component of the magnetic field and the Doppler velocity by using
 the NFI data sets.
 We also used the Spectropolarimeter (SP) which takes the Stokes profile of the two FeI lines at 6301.5 $\AA$ and 6302.5 $\AA$.
 These lines are formed in the photosphere.
 We calculate various physical quantities by fitting the spectra on the assumption of the Milne-Eddington atmospheric model
 (e.g. del Toro Iniesta, 2003; T. Yokoyama et al., 2010, in preparation).
 The field of view is 82 arcsec $\times$ 82 arcsec.
 The number of observational points is 112 in wavelength with resolution of 0.03 $\AA$.
 We use the fast mode SP data.
 The cadence for one step was 3.8 seconds and the duration for one scan was 15 minutes.
 The time duration of SP data is 20 minutes and sometimes 50 minutes due to the synoptic observation of Hinode.
 The time period of NIF observation is from 15:04 UT to 23:30 UT on 2007 October 31th.
 The spatial scale is 0.32 arcsec ${\rm pix}^{-1}$.

 We investigate the temporal evolution of magnetic field from FG by calculating the $CP$ as an indicator of the LOS magnetic
 field.
 $CP$ is defined as 
 
\begin{equation}
  CP = V_b - V_r,
\end{equation}
 Here, $V_b$ is the Stokes-V signal in the blue wing and $V_r$ is the Stokes-V signal in the red wing.
 As the LOS magnetic field becomes stronger, the amplitude of Stokes-V signal arises and then the absolute value of
 $CP$ becomes larger.
 We employ $D$ calculated from the FG data as an indicator of a LOS velocity.
 $D$ is defined as
 
\begin{equation}
  D = \frac{I_b - I_r}{I_b + I_r},
\end{equation}
 $I_b$ and $I_r$ are intensities of blue and red component in Stokes-I signal, respectively.
 The actual zero point of $D$ is set as the average of $D$ over the field of view is zero. 
 We derive the theoretical relationship between $D$ and the actual Doppler-shift using the line profile of solar spectrum atlas 
 and the trancemittance profile of the NFI.
 The relationship between an LOS velocity and $D$ is nearly linear in the range between $-6.0 \ {\rm km}\ {\rm s}^{-1}$
 and $6.0 \ {\rm km}\ {\rm s}^{-1}$. 
 We used 10 as the conversion factor from $D$ to actual Doppler velocity.

\section{Results}

Figure 1 shows the LOS component of magnetic fields taken by SP(left) and the NFI(right).
Left panel is a sub-region of the SP map. The square in the left panel shows the field of view of the right panels.
The plotted time in the left panel is when the slit is at the cancellation site, namely at -547.82 arcsec in x coordination.
Figure 2 shows the scanned maps taken by SOT/SP.
The positive and negative patches with the mean 
magnetic flux density of $\approx 200\ {\rm Gauss}$ were located 
with separation by 2 arcsec at 15:46 UT (left panels in Fig. 2).
These patches approached by apparent proper velocity with 
$\approx 0.2\ {\rm km}\ {\rm s}^{-1}$ and began canceling with each other 
(middle and right panels in Fig. 2).
At 17:21 UT (see Fig. 1), the negative patch 
disappeared completely, while the positive one remained partially. 
The duration of this event defined by the contact of the magnetic patches
and the disappearance of the negative polarity was $\approx 1\ {\rm hour}$
from 16:25 UT to 17:21 UT.

As shown in the bottom panels of Figure 2, it is found that a transverse 
field appears between the patches in the SOT/SP scan maps
during the cancellation process at 16:36 UT and 16:56 UT.
The magnetic flux density of this transverse field 
was $\approx 200\ {\rm Gauss}$ and the direction 
was in parallel with that connecting 
the magnetic patches. Note that because of the azimuth-angle 
180 degree ambiguity, we do not know, at this moment, whether
the orientation is from positive to negative or opposite.

Figure 3 shows the Dopplergram obtained from the $D$ parameters taken by the SOT/FG observations. Since we are
focusing on the velocity structure corresponding to the
cancellation event with a time scale of $\approx 1\ {\rm hour}$,
the Dopplergrams are temporally averaged by 
using several nearby flames in $\approx 5\ {\rm minutes}$ 
to remove the shorter-period features such 
as the 5-minute oscillations and the granular flows.
From Figure 3, it is found that there is a strong red-shift (downflow) signature around/at the canceling magnetic patches.
This red-shift appeared with $\approx 1.3 \ {\rm km}\ {\rm s}^{-1}$ from 16:43 UT to 17:25 UT until the end
 of the canceling process and significantly long-lasting compared to the 5-minutes oscillation (see Fig.4).

Figure 4 shows the temporal evolution of the spatially integrated $CP$ parameter as a proxy of the magnetic flux
 for each of the magnetic patches.
The magnetic flux of the negative patch monotonically decreased from the
beginning of the cancellation at 16:25 UT to the end at 17:21 UT.
Except for the temporary increase in 16:45 -- 17:00 UT
caused by a coalescence of two adjacent positive patches, the positive
flux was also decreasing by the similar rate with the negative flux
in the same period. The time of the observed transverse field connecting the canceling patches 
are indicated by the arrows at 16:36 UT and 16:56 UT, which are well in the period
of the canceling processes.
 Each asterisk indicates the averaged red-shift speed in the $5\ pixel \times 5\ pixel$ box centered by the peak red-shift 
nearby the canceling patches at each time. 
 This speed is averaged over 25 pixels around the max velocity point.
 The red-shift began to increase from 16:43 and reached $\thicksim 1.7\ {\rm km}\ {\rm s}^{-1}$ at 16:55 UT.
 It decreased rapidly after the cancellation at 17:21 UT. 
 The averaged value is $\approx 1.0\ {\rm km}\ {\rm s}^{-1}$ in this period during the cancellation.
Note that there was a delay by 20 min for this red-shift signature against the start of the cancellation.
 
\section{Discussions}

We studied detailed magnetic and velocity structures of
a cancellation event by using the unprecedented high quality
data obtained by the SOT on board Hinode.
Our observations are summarized as follows: 
(1) A transverse field connecting the pair magnetic patches is found
during the canceling process. 
(2) There was a long-lasting
red-shift (downflow) signature with  
$\approx 1.3 \ {\rm km}\ {\rm s}^{-1}$
around the canceling magnetic patches lasting for 40 minutes.
The start of this Doppler red-shift delays by 20 minutes from the start
of the canceling process.

The transverse field connecting the canceling patches
are consistent with both the ``U-loop emergence''
and ``$\Omega$-loop submergence'' scenarios.
It is still difficult to discriminate the preferable scenario 
only from the magnetic information
since we could not identify the orientation of the 
transverse field due to the azimuth-angle 
180 degree ambiguity. Note if
the orientation is from positive patch to negative, it is consistent
with the ``$\Omega$-loop''  and the vice versa. 
The velocity information in our data sets help the interpretation
as shown in the following discussion.

There was a steady red-shift signature with  
$\approx 1.0 \ {\rm km}\ {\rm s}^{-1}$
around the canceling magnetic patches for 40 minutes.
We can calculate the total amount of the disappeared vertical magnetic flux ($\Phi_{ver}$) by the cancellation and that
 of the submerged horizontal flux ($\Phi_{hor}$) by assuming the direction of the transverse field and
 velocity field.
These fluxes should be comparable.
We assume that the direction of the transverse field is connecting from the positive patch to the negative patch
 in the $\Omega$-loop scenario. 
 The amount of $\Phi_{ver}$ is equivalent with the
flux in the single isolated patch, namely,
$$
\Phi_{ver}=B_{ver}\ S = 4.9 \times 10^{18}\ {\rm Mx}
$$
where $B_{ver} \approx B_{LOS} \ \cos \theta \approx 180 \ {\rm Gauss}$ and 
$S \approx 5$ ${\rm arcsec}^2 \approx 2.7 \times 10^{16}$ ${\rm cm}^2$ are 
the magnetic flux density and the area 
of the negative patch before the
cancellation process, respectively (left panel of Fig. 2).
The amount of the horizontal field in the submerging flux is
$$
\Phi_{hor}=B_{hor}\ w v t = 5.7 \times 10^{18}\ {\rm Mx}
$$
where $B_{hor} \approx B_{trans} \ \cos \theta \approx 270\ {\rm Gauss}$ and $w \approx 1\ {\rm arcsec}
\approx 730\ {\rm km}$ are
the magnetic flux density and the width of the 
submerging horizontal component (middle panel of Fig. 2), respectively. 
$v \approx 1.2\ {\rm km}\ {\rm s}^{-1}$ is the
submerging speed calculated from red-shift $\approx 1.0km/s$ and $t \approx 40\ {\rm min}$ is its lifetime.
Since $\Phi_{ver}$ and $\Phi_{hor}$ have similar values,
the submergence scenario can explain the local budget of the magnetic flux.

What is the submerging mechanism of the $\Omega$-loop submergence?
From the dynamical point of view, the magnetic buoyancy might prevent the $\Omega$-loop from submerging.
To overcome this buoyancy, the magnetic tension force has to be 
stronger, i.e. the curvature radius of the $\Omega$-loop has to be short
enough. The critical value is around several times of the pressure scale height.
The observed distance between the pair patches as a proxy of the curvature radius was
$\approx 1 \ {\rm arcsec}=730\ {\rm km}$ (Fig. 2) which was three times
the photospheric pressure scale height ($\approx 250\ {\rm km}$) and
was short enough for the excess of the tension over the buoyancy force.
The red-shift speed is a fraction ($\approx 10\ \%$) of the estimated
local Alfv\'en speed $C_A \approx 14\ {\rm km}\ {\rm s}^{-1}$ 
by using the observed $B_\perp=200\ {\rm Gauss}$ and assuming the photospheric
density $n=10^{17} \ {\rm cm}^{-3}$. This is again consistent with the
submergence by the magnetic tension force.

We also found the delay of the red-shift speed relative to the cancellation.
We do not understand the reason yet but we speculate that this delay may show the necessity of the convergence of
 the magentic patches which is enough for the magnetic tension to overcome the magnetic buoyancy.
Figure 5 shows our speculative model for a quiet-sun cancellation based on the ``$\Omega$-loop submergence''
scenario.
First, a pair of isolated magnetic fluxes are convected by the
supergranulation flows. When they come in contact at a short distance,
they reconnect at a certain height (not specified here) above the Na I forming layer.
to generate the horizontal field, i.e. an ``$\Omega$-loop'' structure. 
The foot-points of the magnetic fluxes meanwhile continue to be passively convected by the flows to converge
 until the magnetic tension overcomes the magnetic buoyancy. 
Finally, when the magnetic fluxes become close enough with each other,
 the magnetic tension dominates over the magnetic buoyancy.
 Then, the ``$\Omega$-loop'' begins to submerge.
 In this scenario, the main cause of the submergence is the tension of the magnetic flux.
Note that Cheung et al.(2008) refers that the $\Omega$-loop submergence caused by the magnetic tension in their numerical
 simulation and the observed downflow around the cancellation site of the flux emergence region.

Our submerging speed ($\approx 1.2 \ {\rm km}\ {\rm s}^{-1}$) had 
the similar value with that ($\approx 1 \ {\rm km}\ {\rm s}^{-1}$)
reported by Chae et al. (2004) in an active-region cancellation.
It is also consistent with the statistical study by Harvey et al. (1999) 
with the apparent speed ($\approx 1 \ {\rm km}\ {\rm s}^{-1}$)
obtained by using the temporal difference between the timings of the cancellation process in the chromospheric
and in the photospheric levels.
In numerical simulation, Cheung et al.(2008) reports much larger submerging speed.
The downflows are larger in both numerical simulation ($\approx 6\ {\rm km} \ {\rm s}^{-1}$) and observation
 ($\approx 10\ {\rm km} \ {\rm s}^{-1}$) than that in this study.
Such a super-sonic flow may be caused by the small length of the loop in their numerical simulation $\approx 0.3$
 arcsec and by the strong surrounding magnetic loops.

Kubo \& Shimizu (2007) reported that there are both upflow and downflow around the cancellation sites and the direction
of the horizontal field is nearly aligned to the polarity inversion line between the canceling pair in active regions.
They interpreted the high Doppler velocity with both blue and red component as the velocity of the material flow along
 the magnetic field from the center-to-limb variation of the Doppler velocity.
Their cancellations are frequently associated by a filament formation that might suggest an emergence
of a new flux at the cancellation site. This suggests that the mechanism of the cancellation might be different between the
active regions and quiet sun. 
Note that there is a possibility that the observed red-shift is the velocity of the material flow along the field line.
It requires the statistical study of the quiet Sun cancellations to examine this possibility. 
If our velocity structure around the cancellation site is magnetic flux motion, we would not see
 the centor-to-limb variation in the Doppler velocity.

It is necessary to check the generality of our scenario in the application for the cancellation events ubiquitously 
observed in the quiet Sun. We are doing the statistical studies and will report in a separate paper. It is also
interesting to investigate the structures above the cancellation sites in the upper chromosphere, transition region
 and corona.
Since reconnection events in these regions might take place along
with the photospheric cancellation, explosive brightenings
might be observed in UV, EUV or X-rays \citep{bro08}. 
The data sets of the other two Hinode instruments, EIS and XRT, are
under our on-going investigation. 

\acknowledgments

We express our appreciation to my colleague in ISAS/JAXA, NAOJ, Kyoto University, and the University of Tokyo
 for useful discussion. 
We are grateful to Drs. M. Kubo and Y. Katsukawa for useful comments.
Hinode is a Japanese mission developed and launched by ISAS/JAXA, with NAOJ as domestic partner and NASA and STFC (UK) as
 international partners. It is operated by these agencies in co-operation with ESA and NSC (Norway). 
This research made use of data obtained from Data ARchives and Transmission System (DARTS), provided by Center for
 Science-satellite Operation and Data Archives (C-SODA) at ISAS/JAXA.
This work was partly carried out at the NAOJ Hinode Science Center, which is supported by the Grant-in-Aid for Creative
 Scientific Research "The Basic Study of Space Weather Prediction" from MEXT, Japan (Head Investigator: K. Shibata), generous
 donations from Sun Microsystems, and NAOJ internal funding.

\clearpage

\begin{figure}
\epsscale{1.0}
\plotone{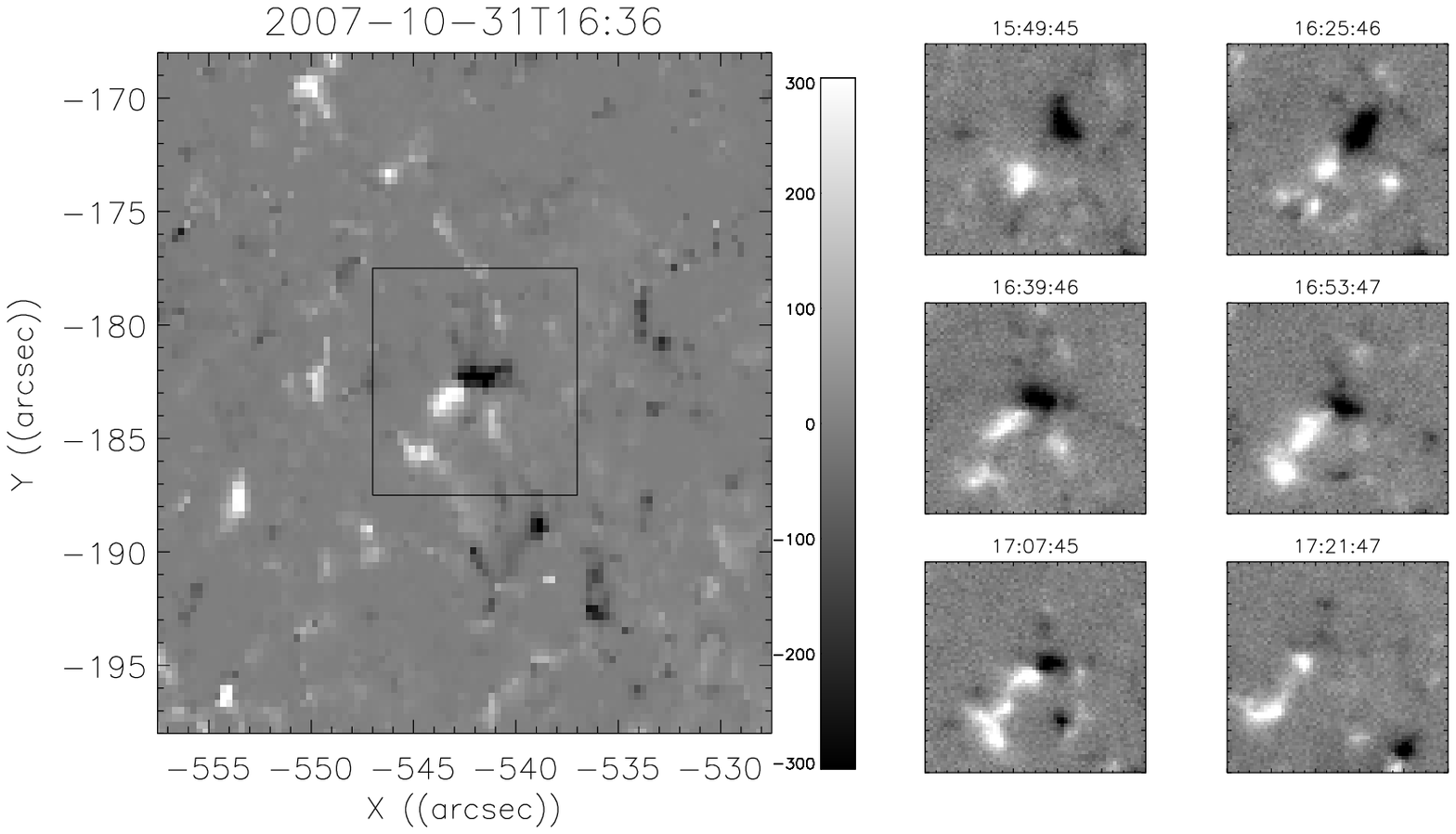}
\caption{LOS magnetic field obtained by SOT/SP (left) and SOT/FG (right).
 The square in the left panel shows the field of view of right panels.
 The time on the left panel is when the slit is at the cancellation site.
 The fields of view of the right panels are $ 10\ {\rm arcsec} \times 10\ {\rm arcsec} $.
}
\label{f1}
\end{figure}

\begin{figure}
\epsscale{1.0}
\plotone{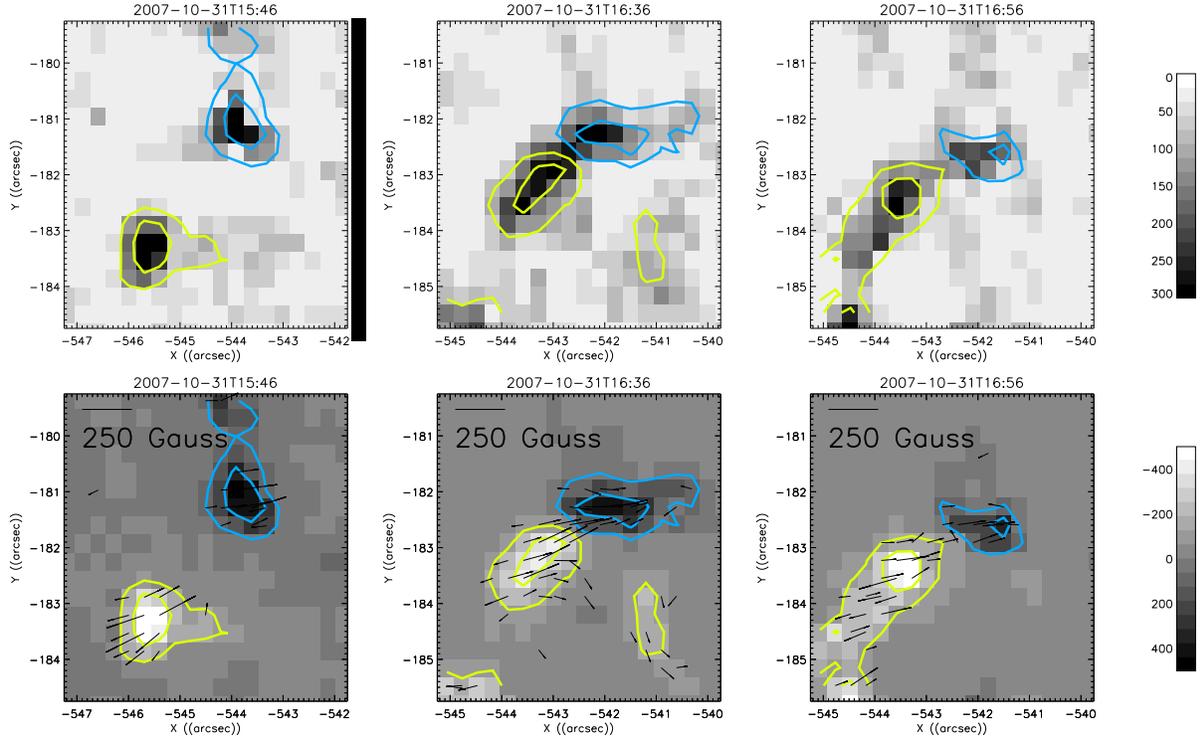}
\caption{The strength and direction of the transverse magnetic field in unit of Gauss obtained from SOT/SP.
 (Top) The background shows the strength of the transverse field and the yellow/blue contours are positive/negative 
LOS magnetic field. The counters show $\pm$100Gauss and $\pm$300Gauss.
 (Bottom) The white/black background shows positive/negative LOS magnetic field. 
 The arrows show the direction of the transvers field.
 The contours show the same strength as the top panels.
 Note that there is a 180-degree ambiguity in the transverse field direction.}
\label{f2}
\end{figure}

\begin{figure}
\epsscale{0.99}
\plotone{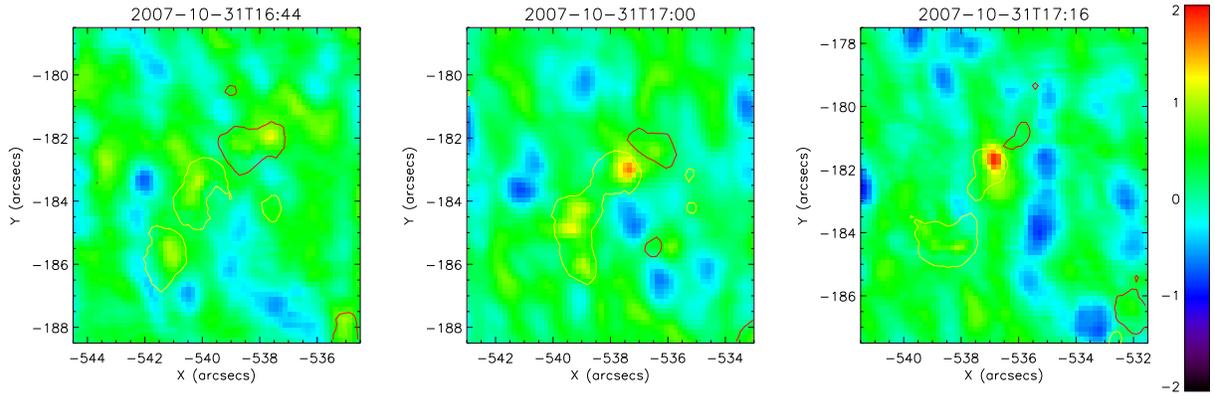}
\caption{The background shows LOS Doppler velocity fields calculated from SOT/FG and the yellow/red contours show
 the positive/negative $CP$ value as an indicator of the LOS magnetic field.
 The plotted LOS velocity range is between -2.0 km/s and 2.0 km/s.}
\label{f3}
\end{figure}

\begin{figure}
\epsscale{0.99}
\plotone{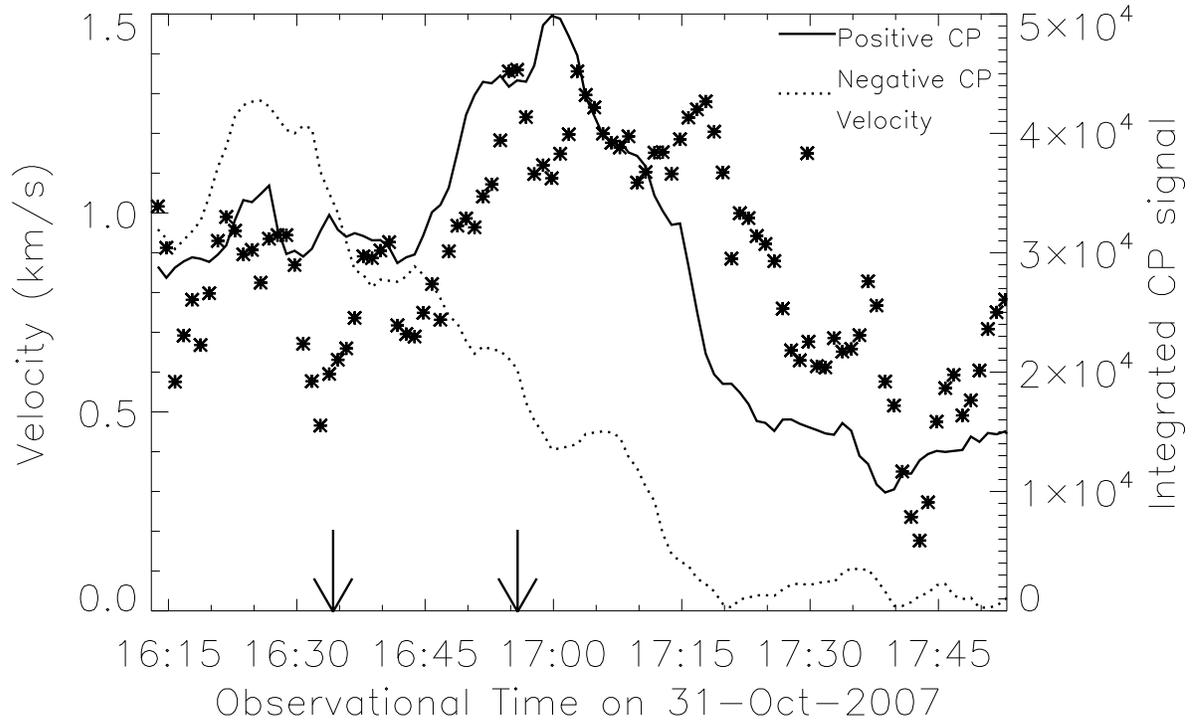}
\caption{The time series of positive $CP$ signal, negative $CP$ signal as an indicators of magnetic fluxes, and the
 Doppler shift around the cancellation site.
 The solid (dotted) line is the positive (negative) $CP$ signals and the asterisks show the Doppler shift averaged
 over the canceling site.
 The arrows indicate the timing of the SP observations which include the transverse field connecting the canceling patches.}
\label{f4}
\end{figure}

\begin{figure}
\epsscale{1.0}
\plotone{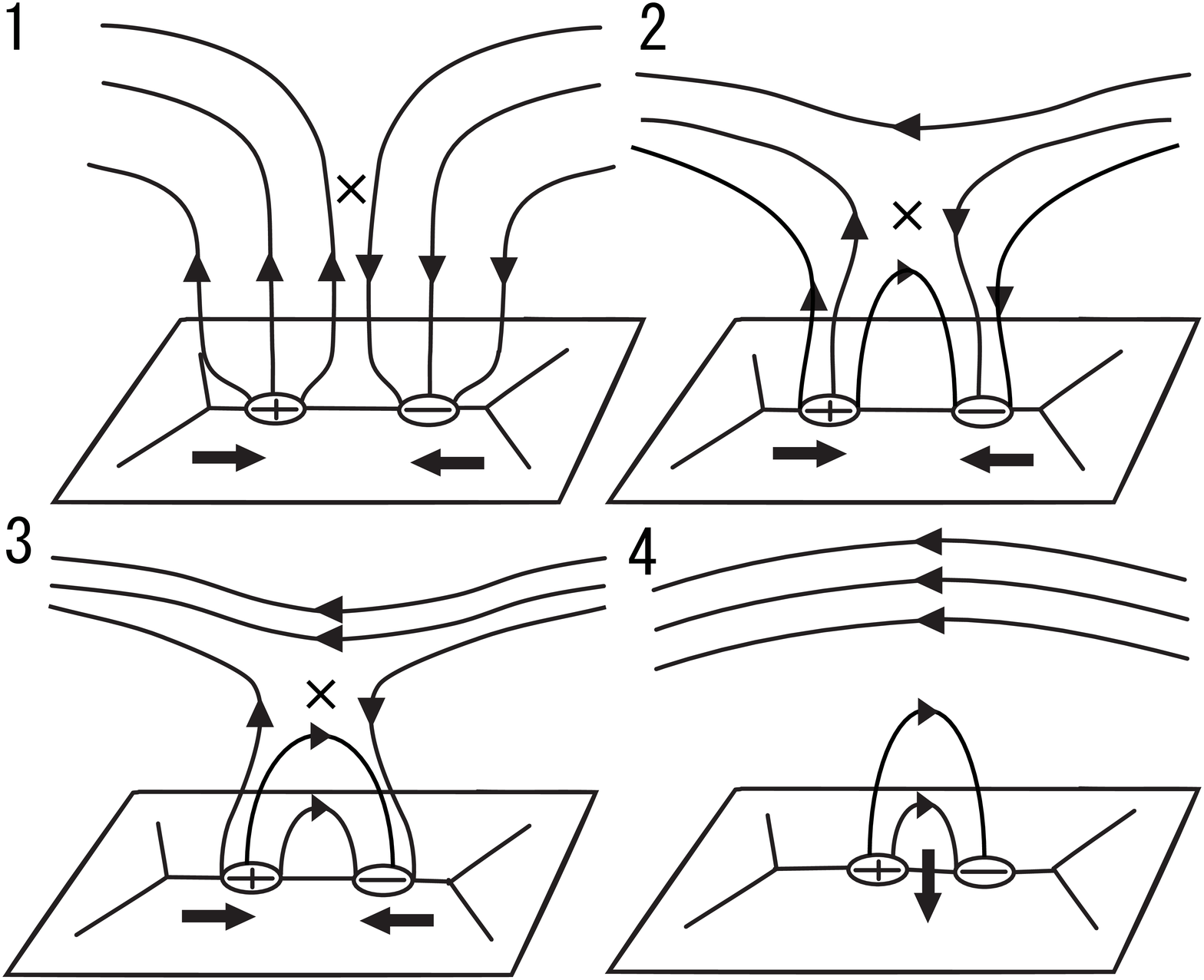}
\caption{A speculative picture of an $\Omega$-loop submergence model.
 First, two anti-parallel magnetic fluxes are converging toward each other (Panel 1).
Then magnetic reconnection occurs above the photosphere and the fluxes continue to converge (Panel 2 and 3). When the magnetic tension becomes
 stronger than the magnetic buoyancy due to the convergence, the magnetic loop begins to submerge (Panel 4).}
\label{f5}
\end{figure}

\end{document}